\documentclass[%
 reprint,
superscriptaddress,
 amsmath,amssymb,
 aps,
]{revtex4-2}

\usepackage{graphicx}
\usepackage{bm}
\usepackage{soul}
\usepackage[colorlinks=true,allcolors=blue]{hyperref}
\usepackage{xcolor}

\newcommand{\M}{\mathsf{M}}
\newcommand{\mS}{\mathcal{A}}
\DeclareMathOperator{\Tr}{Tr} 

\begin{document}

\preprint{APS/123-QED}

\title{
Information-optimal mixing at low Reynolds number}

\author{Luca Cocconi}
\email{luca.cocconi@ds.mpg.de}
\affiliation{%
 Max Planck Institute for Dynamics and Self-Organization, G{\"o}ttingen 37073, Germany
}%
\author{Yihong Shi}%
\affiliation{%
 Max Planck Institute for Dynamics and Self-Organization, G{\"o}ttingen 37073, Germany
}%
\author{Andrej Vilfan}%
\affiliation{Jožef Stefan Institute, 1000 Ljubljana, Slovenia}

\date{\today}

\begin{abstract}
Mutual information between particle positions before and after mixing provides a universal assumption-free measure of mixing efficiency at low Reynolds number which accounts for the kinematic reversibility of the Stokes equation. For a generic planar shear flow with time-dependent shear rate, we derive a compact expression for the mutual information as a nonlinear functional of the shearing protocol and solve the associated extremisation problem exactly to determine the optimal control under both linear and non-linear constraints, specifically total shear and total dissipation per unit volume. Remarkably, optimal protocols turn out to be universal and time-reversal symmetric in both cases. Our results establish a minimum energetic cost of erasing information in a broad class of non-equilibrium drift-diffusive systems.

\end{abstract}

\maketitle

Because of the kinematic reversibility of the Stokes equation \cite{Purcell1977,arrieta2020geometric}, most compellingly illustrated by G. I. Taylor's Couette cell experiment \cite{national1972illustrated,Heller1960}, fluid mixing at low Reynolds number requires an interplay between advection (stirring) and diffusion \cite{villermaux2019mixing,tang2020quantifying}. Shear-induced enhancement of diffusive mixing, a phenomenon closely related to Taylor dispersion \cite{taylor1953dispersion}, is fundamental to many biological and artificial systems, from the uptake
of oxygen, nutrients, or chemical signals in ciliated aquatic microorganisms to microreactors and “lab-on-a-chip” applications \cite{stroock2002chaotic,campbell2004microfluidic,grigoriev2006chaotic,pine2005chaos,aref2017frontiers}. In fact, it represents a fundamental feature of any out-of-equilibrium relaxation process governed by an advection-diffusion equation \cite{villermaux2019mixing}, including the dispersion of pollutants in the  upper  troposphere and stratosphere \cite{konopka1995analytical}.
As a result, the design
of optimal mixing protocols is a problem of both fundamental
and practical importance \cite{d1999control,shi2024mutual,LIN_THIFFEAULT_DOERING_2011,GUBANOV_CORTELEZZI_2010} and aligns with a growing interest in applying concepts of optimal control theory to nonequilibrium physics \cite{schmiedl2007optimal,loos2024universal,proesmans2020optimal,garcia2024optimal,cocconi2024dissipation,engel_control2023,davis_control2024,Proesmans2023}. 

Traditionally, global mixing efficiency has been quantified by imposing an initial pattern (e.g.\ solute distribution or temperature profile) and characterising the effect of stirring on the latter through the change in its $L^2$/Sobolev norms \cite{danckwerts1952definition,thiffeault2012using} or Shannon entropy \cite{d1999control,camesasca2006quantifying,thiffeault_nonuniform2021}. Local mixing may additionally be quantified in terms of Lyapunov exponents \cite{arrieta2020geometric,tsang2005exponential}.
More recently, a universal assumption-free (i.e. pattern-independent) metric for global mixing efficiency was introduced in the form of the mutual information between particle positions before and after mixing \cite{shi2024mutual}. In experiments, mutual information can be estimated from tracer data using lossless compression algorithms \cite{de2022entropy}.

Here, we apply this novel metric to the problem of mixing of a fluid by a divergence-free linear shear flow. Defining the time-dependent shear rate as our protocol, we re-express the mutual information as a non-linear functional of the latter and solve the optimal control problem exactly to derive optimal protocols under constraints of total shear and total viscous dissipation per unit volume.
From these, we derive exact bounds on the mixing efficiency.
Our findings demonstrate that mutual information is not only a conceptually elegant framework, but also that its analytical tractability may offer practical advantages in the determination of optimal mixing protocols.\\

\begin{figure}
    \centering
    \includegraphics[width=0.9\columnwidth]{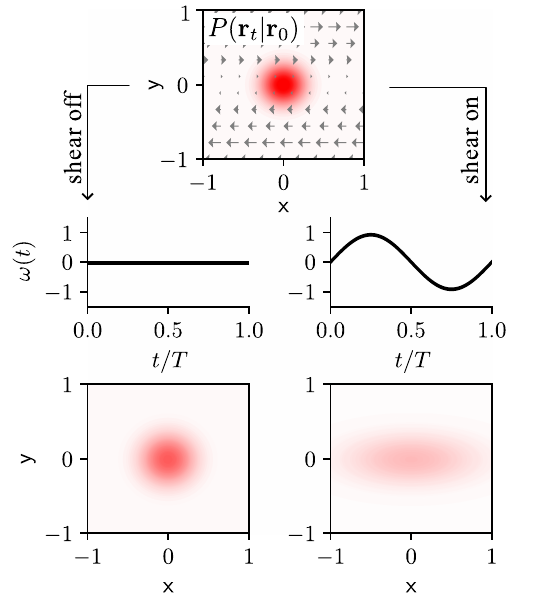}
    \caption{Enhancement of diffusive mixing in the presence of a time-dependent divergence-free flow (Taylor dispersion), shown here for the case of simple shear, $\mathbf{v}(\mathbf{r},t) = \omega(t)(y,0)$.  
    }
    \label{fig:introfigure}
\end{figure}

\paragraph*{General framework ---}Let $\mathbf{r}(t) \in \mathbb{R}^d$ denote position in a fluid and $\mathbf{v}(\mathbf{r},t)$ be a linear shear flow with time-dependent shear rate $\omega(t)$ put under the control of an external operator, $v_i(\mathbf{r},t)=\omega(t)\M_{ij} r_j$ (Fig.~\ref{fig:introfigure}).
We assume that the flow is divergence-free, $\nabla\cdot\mathbf{v}=0$, implying $\Tr(\M)=0$.
Following the approach introduced in Ref.~\cite{shi2024mutual}, we quantify the mixing efficiency of a \emph{protocol} $\{\omega(t)\}_0^T$ of duration $T$ in terms of (minus) the mutual information $I[\mathbf{r}_t;\mathbf{r}_0]$ between the initial and final position of a tagged point particle that is perfectly advected by the flow and simultaneously undergoes Brownian motion with constant diffusivity $D$.
The mutual information can be written in terms of the Gibbs-Shannon entropy $S$ as
\begin{equation} \label{eq:mutual_info_h}
        I[\mathbf{r}_T;\mathbf{r}_0] = S[P(\mathbf{r}_T)] - \int d\mathbf{r}_0 P(\mathbf{r}_0) S[P(\mathbf{r}_T|\mathbf{r}_0)]
    \end{equation}
where $P(\mathbf{r}_T)$, $P(\mathbf{r}_0)$ and $P(\mathbf{r}_T|\mathbf{r}_0)$ denote the posterior, prior and conditional probability densities of the process. 
For the natural assumption-free choice of uniform prior distribution, the divergence-free condition implies a uniform posterior. Thus, only the conditional probability depends on the shearing protocol.
In particular, the conditional density is governed by the advection-diffusion equation $\partial_t P(\mathbf{r}_t|\mathbf{r}_0) + \bm{\nabla}\cdot(\mathbf{v}P - D\bm{\nabla} P)=0$ and may describe the time-dependent solute/temperature distribution given a Dirac-delta initialisation $\delta(\mathbf{r}-\mathbf{r}_0)$ at $t=0$.
Equivalently, it captures the time-dependent distribution of a tagged particle whose position $\mathbf{r}(t)$ is governed by the linear Langevin dynamics 
\begin{equation}\label{eq:langevin_bw}
    \dot{\mathbf{r}}(t) = \omega(t)\M \mathbf{r}(t) + \sqrt{2D} \bm{\eta}(t)
\end{equation}
with Gaussian white noises acting on each coordinate, $\langle \eta_i(t_1)\eta_j(t_2)\rangle =\delta_{ij}\delta(t_1-t_2)$. 
The conditional probability density associated with Eq.~\eqref{eq:langevin_bw} is thus a multivariate Gaussian of the form
\begin{equation}\label{eq:trivariate_gauss}
        P(\mathbf{r}_t |\mathbf{r}_0) = \frac{1}{\sqrt{(2 \pi )^d \det\Sigma_t }} e^{-\hat{\mathbf{r}}_t^T \Sigma_t^{-1} \hat{\mathbf{r}}_t} 
\end{equation}
where $\hat{\mathbf{r}}_t = \mathbf{r}_t - \overline{\mathbf{r}}_t$ denotes the deviation from the mean position, which is obtained by solving Eq.~\eqref{eq:langevin_bw} with $D=0$,
and $\Sigma(t)$ is the covariance matrix, which is independent of the initial position $\mathbf{r}_0$. Using the standard result for the entropy of a multivariate Gaussian distribution we find
\begin{equation}\label{eq:h_normal}
    S[P(\mathbf{r}_t|\mathbf{r}_0)] = \frac{d}{2}\ln(2\pi e) + \frac{1}{2}\ln \det\Sigma(t)\;,
\end{equation}
which is again independent of $\mathbf{r}_0$. Substituting into Eq.~\eqref{eq:mutual_info_h}, performing the now trivial integral over $\mathbf{r}_0$ and recalling that $S[P(\mathbf{r}_t)]$ is independent of $\omega(t)$, 
we conclude that maximum mixing efficiency is achieved by protocols that maximise the covariance determinant.
Noting that $\Sigma_{ij}(0)=0$, the covariance itself may be written explicitly as \cite{van1976expansion}
\begin{equation}\label{eq:Sigma_expproduct}
    \frac{\Sigma(t)}{2D} = \int_0^t dt_1 \exp\left[ \M\int_{t_1}^t dt_2 \omega(t_2) \right] \exp\left[ \M^T \int_{t_1}^t dt_2 \omega(t_2) \right]\,.
\end{equation}
Although the above equation provides a formal solution for $\det \Sigma$, its calculation can be simplified by trading the determinant for another time integral. We combine the Jacobi formula for the determinant of a generic time-dependent matrix $\Sigma(t)$
\begin{equation}
        \frac{d}{dt} \det\Sigma(t) = \det\Sigma(t) \cdot \Tr\left( \Sigma^{-1}(t) \frac{d\Sigma(t)}{dt}\right)
\end{equation}
with the Lyapunov equation 
that follows from the time derivative of Eq.~(\ref{eq:Sigma_expproduct})
\begin{equation}
        \frac{d\Sigma(t)}{dt} = \left[ \M \Sigma(t) + \Sigma(t) \M^T \right]\omega(t) + 2D \mathbb{I}
\end{equation}
where $\mathbb{I}$ is the identity matrix,
to obtain 
\begin{equation}\label{eq:dtSigma}
    \det\Sigma(t) = 2D  \int_0^t dt_1 \ \det\Sigma(t_1) \Tr\Sigma^{-1}(t_1)\;.
\end{equation}
Eq.~\eqref{eq:dtSigma} draws solely on $\M$ being traceless, a property which is inherited from the requirement that $\mathbf{v}(\mathbf{r},t)$ is a divergence-free flow. \\

In the following, we restrict ourselves to the case of a 2D fluid ($d=2$), whereby both $\Sigma$ and $\M$ are $2\times 2$ matrices. Our results will also apply to 3D fluids upon assuming translational invariance along one of the dimensions \footnote{In this case, one of the three coordinates (say, $z(t)$) reduces to an independent Wiener process and the covariance matrix $\Sigma$ acquires a block diagonal structure. Its determinant is given by the product of the nontrivial upper block, which accounts for the coupled dynamics in the $(x,y)$ plane and is identical to the covariance for the 2D case, and the $\Sigma_{zz}(t)=2Dt$ element, which is protocol independent.}. We may now write $\M$ as a linear superposition:
\begin{equation}\label{eq:M_lineardecomp}
    \M =
    m_r
    \begin{bmatrix}
           0 & 1 \\
           -1 & 0
    \end{bmatrix}    
    +
    \frac{m_{p,1}}{2}
    \begin{bmatrix}
           0 & 1 \\
           1 & 0
    \end{bmatrix}    
    +
    \frac{m_{p,2}}{2}
    \begin{bmatrix}
           1 & 0 \\
           0 & -1
    \end{bmatrix}    \;,
\end{equation}
where the subscripts $r,p$ refer to rotation and pure shear, respectively. Simple shear along the $x$ axis, for example,  corresponds to $2 m_r = m_{p,1}$ with $m_{p,2}=0$.
For $2\times 2$ matrices, Eq.~\eqref{eq:dtSigma} reduces to
\begin{equation}\label{eq:det_tr_relation_2x2}
    \det\Sigma(t) = 2D  \int_0^t dt_1  \Tr\Sigma(t_1)\;,
\end{equation}
where the integrand no longer contains the covariance determinant. $\Tr \Sigma$ is determined from Eq.~\eqref{eq:Sigma_expproduct}, which can be massaged into a more convenient form by evaluating the traceless matrix exponentials using Putzer's method \cite{putzer1966avoiding}
\begin{equation}\label{eq:cases_exp_matrix}
    e^{k \M} = 
    \begin{cases}
        \cosh(\lambda_\M k) \mathbb{I} + \frac{\sinh(\lambda_\M k)}{\lambda_\M} \M \quad &\text{if  $\lambda_\M  \neq 0$} \\
         \mathbb{I} + k\M \quad &\text{if $\lambda_\M =0$}  
    \end{cases}
\end{equation}
where $k \in \mathbb{R}$. The eigenvalue $\lambda_\M = \sqrt{-\det\M}$ is non-negative real when $m_{p,1}^2 + m_{p,2}^2 \geq 4m_r^2 $ and purely imaginary otherwise. Notice that the case $\lambda_\M =0$ (e.g.\ simple shear) can be recovered in the limit $\lambda_\M \to 0$ and thus need not be treated separately. 
For $\lambda_\M \neq 0$, and using $e^{k\M^T} = (e^{k\M})^T$, we obtain 
\begin{align}
         \Tr\Sigma = 2D \int_0^t & dt_1 \ 2 \cosh^2\left(\lambda_\M \int_{t_1}^t dt_2 \omega(t_2) \right) \nonumber \\
         + &\frac{\sinh^2\left(\lambda_\M \int_{t_1}^t dt_2 \omega(t_2) \right)}{\lambda_\M^2} \Tr(\M\M^T)\;.
\end{align}
Equation \eqref{eq:det_tr_relation_2x2} then leads to 
\begin{align}
         \det\Sigma(T)
         &= 4 D^2 T^2 + 2D^2 \mS[\omega]
\end{align}
where we used the relation $\cosh^2 = 1 +\sinh^2$ and defined a non-negative action-like functional $\mS[\omega]$ as
\begin{align}\label{eq:s_minimal}
        \mS[\omega] &\equiv
        \frac{\gamma_\M^2}{\lambda_\M^2} \int_0^T dt_1 \int_0^{T} dt_2 \sinh^2\left( \lambda_\M \int_{t_1}^{t_2} dt \omega(t)\right)\;.
\end{align}
with $\gamma_\M^2 =  2\lambda_\M^2 + (\M\colon\!\M)$ and $\M\colon\!\M \equiv \Tr(\M \M^T)$. 
This functional can be related back to the ``excess reduction'' in mutual information (equivalently, the increase in mixing efficiency) induced by shearing with respect to a purely diffusive baseline via the  expression
\begin{equation}\label{eq:inf_diff}
    \Delta I \equiv I[\mathbf{r}_T;\mathbf{r}_0]_{\rm noshear} - I[\mathbf{r}_T;\mathbf{r}_0] = \frac{1}{2}\ln\left( 1 + \frac{\mS[\omega]}{2T^2}\right)\;.
\end{equation}
Perhaps counterintuitively, $\Delta I$ does not depend on the diffusivity $D$.
Eqs.~\eqref{eq:s_minimal} and \eqref{eq:inf_diff} constitute one of our key results as they reduce the optimal mixing problem to the maximisation of a relatively simple functional, which vanishes in the absence of shear. 
Consistently with general arguments from the theory stochastic processes \cite{dieball_coarse2022}, the functional $\mS$ is manifestly invariant under both parity- and time-reversal of the shearing protocol, i.e.\ $$\mS[\omega]=\mS[\mathcal{P}\omega]=\mS[\mathcal{T}\omega] = \mS[\mathcal{P}\mathcal{T}\omega]$$
where $\mathcal{P}\omega(t) \equiv -\omega(t)$ and $\mathcal{T}\omega(t) \equiv \omega(T-t)$ with $t \in [0,T]$. 
This implies that, in the absence of additional constraints breaking these symmetries explicitly, global optima $\omega^*$ maximising $\mS$ must be \emph{at least} 4-fold degenerate, if $\mathcal{T}\omega^*\neq \omega^* \wedge \mathcal{P}\mathcal{T}\omega^*\neq \omega^*$, or 2-fold degenerate, if $\mathcal{T}\omega^*= \omega^* \lor \mathcal{P}\mathcal{T}\omega^*= \omega^*$. The numerical finding in Ref.~\cite{shi2024mutual} that globally optimal protocol tend to be $\mathcal{T}$-symmetric is thus non-trivial. We will return to this point shortly.

We henceforth assume $m_{p,1}^2 + m_{p,2}^2 \geq 4m_r^2 $, such that $\lambda_\M \geq 0$ is real. While some of our results may generalise straightforwardly to the case of imaginary $\lambda_\M$, we argue based on Eq.~\eqref{eq:s_minimal} that such protocols would in general offer a comparatively poorer performance, due to the boundedness of the ensuing trigonometric functions, and are thus of lesser interest.

\begin{figure}
    \centering
    \includegraphics[width=\columnwidth]{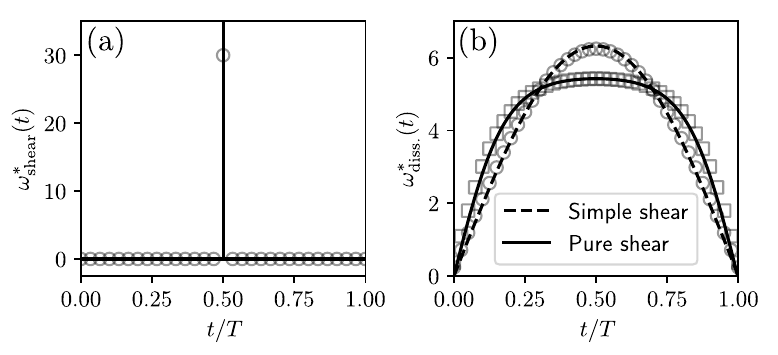}
    \caption{Numerical validation of the optimal protocols for (a) fixed shear (fixing $\Omega=1$) and (b) fixed dissipation (fixing $\sigma T/\eta=5$), showing perfect agreement between theory (lines) and the results of numerical optimisation of the discretised problem (markers). }
    \label{fig:brute_force_check}
\end{figure}

\paragraph*{Optimal protocol under fixed shear ---}
We now proceed to maximise $\mS[\omega]$ as given in Eq.~\eqref{eq:s_minimal} with a linear constraint on total shear
\begin{equation}
  \Omega = \int_0^T dt \gamma_\M \omega(t)
  \label{eq:OmegaDef}
\end{equation}
while demanding $\omega(t) > 0$. 
In the notation of Eq.~\eqref{eq:M_lineardecomp}, the characteristic rate squared $\gamma_\M^2 = m_{p,1}^2 + m_{p,2}^2$ is clearly non-zero for all non-trivial $\M \neq 0$ excluding solid rotation.
By discretisation of the time integrals in Eq.~\eqref{eq:s_minimal} [see Appendix \ref{a:proof_delta}], it can be shown that, for any $\M$, the globally optimal protocol is given by a Dirac-delta impulse at $t=T/2$:
\begin{equation}\label{eq:delta_optimum}
    \omega^*_{\rm shear}(t) = \frac{\Omega}{\gamma_\M} \delta\left(t-\frac{T}{2} \right)\;.
\end{equation}
Figure~\ref{fig:brute_force_check}a shows the result along with the outcome of numerical optimisation. 
Eq.~\eqref{eq:delta_optimum} is qualitatively consistent with the numerical finding of Ref.~\cite{shi2024mutual} that optimal protocols for fixed total shear in a Taylor-Couette cell resemble Dirac delta impulses in the regime $\Omega \ll 1$.
This may be rationalised by noticing that the bulk of a Taylor-Couette flow is well-approximated by homogeneous simple shear for short times and thin gaps.  

In realistic settings, we may want to regularise the optimum \eqref{eq:delta_optimum} through the introduction of a total variation regulator penalising large values of $|\omega'(t)|$, see Appendix \ref{a:regularisation}, where we solve the corresponding problem by means of the Euler-Lagrange equation.

\paragraph*{Optimal protocol under fixed dissipation ---}
In the second scenario, we determine the optimal protocol under a non-linear constraint on total dissipation per unit area
\begin{equation}\label{eq:constraint_dissipation}
    \sigma = \eta \int_0^T dt \sum_{i,j}(\partial_{r_i} v_j + \partial_{r_j} v_i)^2 = \eta \gamma_\M^2 \int_0^T dt \omega^2(t)
\end{equation}
where $\eta$ denotes the dynamic viscosity. 
We now maximise the action $\mS[\omega]$ by demanding that its functional derivative with respect to $\omega(t)$, along with a Lagrange multiplier to enforce the constraint in Eq.~\eqref{eq:constraint_dissipation}, is zero:
\begin{equation}\label{eq:diff_imp}
  \frac{\delta \mS[\omega]}{\delta\omega(t)} - 2 \mu_\sigma \omega(t)=0.
\end{equation}
The derivative follows from Eq.~\eqref{eq:s_minimal} and reads
\begin{equation}\label{eq:S_func_der}
    \frac{\delta \mS[\omega]}{\delta \omega(t)} = 2\frac{\gamma_\M^2}{\lambda_\M^2} \int_0^t dt_1 \int_{t}^T dt_2 \sinh\left( 2 \lambda_\M \int_{t_1}^{t_2} d\tau \omega(\tau)\right)\;.
\end{equation}
It vanishes for $t=0$ and $t=T$, supporting our physical intuition that shear contributes most to the mixing efficiency when performed at intermediate times. 
Differentiating \eqref{eq:diff_imp} thrice with respect to $t$ and using lower order derivatives to substitute the integrals in the resulting expression (see Appendix \ref{a:derivation_ODE} for details), we eventually find that any dependence on the multiplier $\mu_\sigma$ drops out and we are left with the nonlinear ordinary differential equation 
\begin{equation}\label{eq:ode_nonl}
    \omega''(t) = -c^2 \omega(t) + 2 \lambda_\M^2 \omega^3(t)
\end{equation}
with an unknown parameter $c^2>0$. The boundary conditions are natural, $\omega(0)=\omega(T)=0$, as evinced from Eq.~\eqref{eq:diff_imp} by noticing that that $\delta\mS/\delta\omega$ vanishes at the end points [cf.~Eq.~\eqref{eq:S_func_der}].  
Remarkably, Eq.~\eqref{eq:ode_nonl} has a Hamiltonian structure and it is indeed the equation of motion of the anharmonic Duffing oscillator \cite{thompson2002nonlinear}. Its exact solution is given by the Jacobi elliptic functions \cite{salas2021duffing,abramowitz1968handbook}. In particular,
\begin{equation}\label{eq:sn_sol_sigmaparam}
  \omega^*_{\rm diss.}(t) = \frac {2 K(m)\sqrt m}{\lambda_\M T}  {\rm sn}\left( \left. \frac{2K(m)t}{T} \right| m \right)\;,
\end{equation}
where ${\rm sn}$ denotes the elliptic sine and where we chose the constant $c^2$ such that both boundary conditions are satisfied.
After re-parametrising the dissipation $\sigma_\M \equiv \sigma/(\eta \gamma_\M^2)$, we obtain an additional implicit equation for the parameter $0 \leq m<1$ that determines the amplitude, namely $4K(m)(K(m)-E(m)) = \sigma_\M \lambda_\M^2 T$. Here, $K(m)$ and $E(m)$ denote the complete elliptic integral of the first and second kind \cite{abramowitz1968handbook}, respectively.
Since $\lim_{\sigma_\M \lambda_\M^2 T \to 0} m=0$ and ${\rm sn}(u|0)=\sin(u)$, the optimal protocol \eqref{eq:sn_sol_sigmaparam} converges to a sine form, $\omega^*_{\rm diss.}(t)=\sqrt{2\sigma_\M/T}\sin(\pi t/T)$, in the limit of small dissipation or for simple shear, $\lambda_\M=0$. Both results agree with the results of numerical optimisation (Fig.~\ref{fig:brute_force_check}b).

\begin{figure}
    \centering
    \includegraphics[width=0.9\columnwidth]{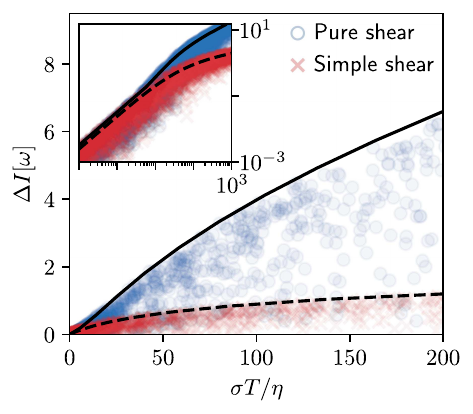}
    \caption{Dissipation bounds on the mixing efficiency. 
    The tight bound (solid line) is obtained by numerical evaluation of Eq.~\eqref{eq:s_minimal} at the optimum, as described in the main text. The dashed line corresponds to the result for simple shear, Eq.~\eqref{eq:bound_Mdep_sigma}.
    Markers indicate the performance of random controls, sampled from independent realisations of a one dimensional Wiener process, $\omega(t)=s W_t$ with $\log_{10}(s) \in (-2,4)$ sampled from a uniform distribution.
    Here, we fix $\lambda_\M/\gamma_\M=1/2$ for pure shear, while for simple shear $\lambda_\M/\gamma_\M=0$.}
    \label{fig:bounds}
\end{figure}

\paragraph*{Bounds on mixing efficiency ---}Having access to explicit expressions for the optimal protocols further allows for the derivation of tight bounds on the mixing efficiency. For the case of fixed shear, 
we substitute \eqref{eq:delta_optimum} into \eqref{eq:s_minimal} and use the relation \eqref{eq:inf_diff} to obtain 
\begin{equation}\label{eq:bounds_shear}
    \Delta I(\Omega;\M) \leq \frac{1}{2}\ln\left[ 1 + \frac{\gamma_\M^2 }{8 \lambda_\M^2} \sinh^2\left(\frac{\Omega\lambda_\M}{\gamma_\M}\right) \right]\;.
\end{equation}
This bound possesses an $\M$-independent quadratic asymptote at small shear, $\Delta I \leq \Omega^2/16 + \mathcal{O}(\Omega^3)$, and is linear at large shear, $\Delta I \leq \Omega\lambda_\M/\gamma_\M + \mathcal{O}(1)$, where it retains an explicit dependence on $\M$.

For the case of fixed dissipation, we may combine Eqs.~\eqref{eq:sn_sol_sigmaparam}, \eqref{eq:s_minimal} and \eqref{eq:inf_diff} to obtain an integral expression for the $\sigma$-dependent bound on the mixing efficiency for a given $\M$.
When $\lambda_\M=0$ (e.g.\ simple shear), this bound is determined by explicitly evaluating Eq.~\eqref{eq:s_minimal}, expanded to lowest nontrivial order in $\lambda_\M$, in the sinusoidal limit discussed above (cf.~Appendix \ref{a:lambda_zero}). 
This produces the expression (Fig.~\ref{fig:bounds}, dashed line)
\begin{align}
    \left. \Delta I(\sigma;\M)\right|_{\lambda_\M=0} \leq \frac{1}{2}\ln \left[ 1 + \frac{\sigma T}{2\eta\pi^2} \right]~.
    \label{eq:bound_Mdep_sigma}
\end{align}
For $\lambda_\M >0$, the action evaluated at the optimum does not have a closed-form expression and the bound has to be determined by numerical integration (Fig.~\ref{fig:bounds}, solid line). The mixing efficiency of the optimal pure shear protocol always surpasses that of simple shear at a given dissipation.
For small $\sigma$, the two bounds approach each other (Fig.~\ref{fig:bounds}, inset), with corrections of order $\mathcal{O}(\lambda_\M^4\sigma_\M^2 T^2)$.  
At larger $\sigma$, we may still construct an explicit, albeit looser, upper bound by evaluating Eq.~\eqref{eq:s_minimal} at constant $\omega(t) = \max_\tau \omega^*_{\rm diss.}(\tau)$. This is discussed in Appendix \ref{a:loose_bound}, where we additionally show that in the large dissipation asymptote $\sigma_\M \lambda_\M^2 T \gg 1$, the mixing efficiency may not grow faster than $\sqrt{\sigma}$.

\paragraph*{Conclusion ---} We have derived exact expressions for the information-optimal mixing protocols associated with a generic planar shear flow with time-dependent shear rate under constraints of total shear [Eq.~\eqref{eq:delta_optimum}] and total dissipation [Eq.~\eqref{eq:sn_sol_sigmaparam}]. 
The generalisation to more complex constraints is straightforward: for example, one could determine the optimal control at fixed dissipation while also requiring the total shear $\Omega$ to be zero, such that the fluid returns to the original position like in Taylor's experiment \cite{national1972illustrated}.
The functional form of the optimal protocols is found to be universal, i.e. independent of the shear matrix $\M$, and to inherit the time-reversal symmetry of the mixing efficiency ($\omega^* = \mathcal{T}\omega^*$). 
These results allowed us to compute explicit bounds on the mixing efficiency, as given in Eq.~\eqref{eq:bounds_shear} for fixed shear and Eq.~\eqref{eq:bound_Mdep_sigma} for fixed dissipation. 
The latter result is of particular significance, as it establishes a universal limit on the degradation/erasure of information in drift-diffusive systems, a principle for non-equilibrium physics that extends beyond the immediate context of low Reynolds number mixing.

Given the highly non-trivial form of the functional \eqref{eq:s_minimal}, which defines the optimal control problem, it is remarkable that such closed-form results could be obtained, pointing to mutual information as a promising avenue for further analytical studies of low-Reynolds mixing and enhancement of out-of-equilibrium relaxation more broadly. 
In particular, future work could explore how the optimal protocols and bounds presented here generalise to non-trivial 3D shear flows or linear flows where each matrix element $\M_{ij}$ varies independently under a set of protocols $\omega_{ij}(t)$.
The study of finite/periodic systems (e.g.\ the Couette annulus from Ref.~\cite{shi2024mutual}) offers another interesting challenge.
Finally, it is worth pointing out that -- while the uniqueness of the optimal protocols studied here combined with the $\mathcal{T}$-invariance of the nonlinear action implies that the optima inherit this symmetry (see Ref.~\cite{loos2024universal} for a related discussion) -- there may in principle be nonlinear control problems for which the $\mathcal{T}$ symmetry is spontaneously broken in some parameter regime. The search for instances of this phenomenon constitutes a fascinating avenue for future work.

\begin{acknowledgements}
L.C. acknowledges support from the Alexander von Humboldt Foundation. 
A.V. acknowledges support from the Slovenian Research and Innovation Agency (Grant No. P1-0099).
\end{acknowledgements}

\bibliography{bibliography}

\section*{End Matters}

\appendix

\section{\boldmath Simple shear and the $\lambda_\M=0$ case}\label{a:lambda_zero}
Eq.~\eqref{eq:cases_exp_matrix} indicates that the case $\lambda_\M=0$ might require special attention. However, we anticipated that the general results derived in the main text apply to this limiting case upon taking the limit $\lambda_\M \to 0$. Indeed, expanding Eq.~\eqref{eq:s_minimal} to leading order in small $\lambda_\M$ we obtain the quadratic functional
\begin{small}
\begin{equation}\label{eq:S_expanded_smalllambda}
    \mS[\omega] \simeq \gamma_\M^2 \int_0^T dt_1 \int_0^T dt_2 \ (T-\max(t_1,t_2) ) \min(t_1,t_2) \omega_{t_1}\omega_{t_2}\;, 
\end{equation}
\end{small}%
Using simple shear as an example, we now show that Eq.~\eqref{eq:S_expanded_smalllambda} is exact in the case $\lambda_\M = 0$.

Consider simple shear along the $x$ coordinate axis, $m_{p,1}=2m_r $ with $m_{p,2}=0$ in Eq.~\eqref{eq:M_lineardecomp}. We thus have $\lambda_\M=0$ and, without loss of generality, we rescale rates such that $2m_r + m_{p,1} =1$, whereby $\gamma_\M  = 1$. Since $y(t)$ reduces to a simple Wiener process with diffusivity $D$, one can directly evaluate the diagonal elements of the covariance matrix:
\begin{align}
    \Sigma_{yy}(t) &= \int_0^t ds_1 \int_0^t ds_2 \langle \dot{y}(s_1)\dot{y}(s_2)\rangle = 2Dt \nonumber \\
    \Sigma_{xx}(t) &=  \int_0^t ds_1 \int_0^t ds_2 \langle \dot{x}(s_1)\dot{x}(s_2)\rangle \nonumber \\
    &= 2D\left[t + \int_0^t ds_1 \int_0^t ds_2 \ \omega(s_1)\omega(s_2) {\rm min}(s_1,s_2)\right]\;.
\end{align}
Using Eq.~\eqref{eq:det_tr_relation_2x2} and dropping terms that are independent of the shearing protocol, we conclude that ${\rm argmin}_\omega \{ I[\mathbf{r}_t;\mathbf{r}_0] \} = {\rm argmax}_\omega \{\mS_{\rm simple}[\omega]\}$ with a quadratic action $\mS_{\rm simple}[\omega]$ given exactly by the right-hand side of Eq.~\eqref{eq:S_expanded_smalllambda}. 

The optimal protocol for fixed total shear $\Omega$ and $\omega>0$ can be determined by identifying the upper bound $\mS_{\rm simple} \leq {\rm max}(V)\Omega^2$ where $V(t_1,t_2) \equiv (T-\max(t_1,t_2) ) \min(t_1,t_2)$ and noticing that the Dirac-delta impulse $\omega^*(t) = \Omega \delta(t-T/2)$ saturates it, in agreement with Eq.~\eqref{eq:delta_optimum}.
The bound itself follows from the possibility to interpret $\omega(t_1)\omega(t_2)/\Omega^2$ as a normalized measure.

The optimum under the constraint of total dissipation $\sigma$ can be determined by augmenting the action $\mS_{\rm simple}$ through a suitable Lagrange multiplier enforcing Eq.~\eqref{eq:constraint_dissipation} and setting its functional derivative to zero to obtain the integral-operator eigenvalue equation
\begin{equation}\label{eq:eigenv_eq_S}
    \int_0^T \min(t_1,t)(T-\max(t_1,t)) \omega^*(t_1) = \mu_\sigma \omega^*(t)\;.
\end{equation}
It follows from substituting \eqref{eq:eigenv_eq_S} into \eqref{eq:S_expanded_smalllambda} that $\mS_{\rm simple} =\mu_\sigma \sigma/\eta$. We are thus specifically after the largest eigenvalue $\mu_\sigma^{(\rm max)}$ of the operator appearing in \eqref{eq:eigenv_eq_S}. Noticing that \eqref{eq:eigenv_eq_S} also entails natural boundary conditions at $t=0$ and $t=T$, the eigenfunctions are $f_n(t)=\sin(n\pi t/T)$ with $n \in \mathbb{N}$ and associated eigenvalues $\lambda_n = T^2/(\pi^2n^2)$. The optimal protocol is thus given by $\omega^*(t) = \sqrt{2\sigma/\eta T} \sin(\pi t/T)$ and it is equal to to the low $\lambda_\M$ limit of the optimal protocol for a generic shear $\M$ derived in the main text. This finding is validated numerically in Fig.~\ref{fig:brute_force_check}.

\section{Proof of Eq.~\eqref{eq:delta_optimum}}\label{a:proof_delta}
Let $\Gamma(t)$ denote the antiderivative of the rescaled shear rate, whereby $\int_{t_1}^{t_2}dt \lambda_\M \omega(t)=\Gamma(t_2)-\Gamma(t_1)$. Applying the constraint of total shear and demanding that $\omega(t)>0$, we have $\Gamma(0)=0$, $\Gamma(T)=\Omega\lambda_\M/\gamma_\M$ and that $\Gamma(t)$ is a monotonically increasing function of $t$.
The right-hand side of Eq.~\eqref{eq:s_minimal} may then be re-written as
\begin{equation}\label{eq:symmetrised_int}
    \mS[\omega] = {\rm const} + \frac{\gamma_\M^2}{2\lambda_\M^2} \int_0^{T} dt_1 \frac{1}{\mathcal{F}(t_1)}\int_0^{T} dt_2 \mathcal{F}(t_2)
\end{equation}
with $\mathcal{F}(t) \equiv e^{2\Gamma(t)}$\;. The discretised form of the integral in Eq.~\eqref{eq:symmetrised_int} reads
\begin{align}
    &\Delta t^2 \left( \mathcal{F}_i + \sum_{m_1 \neq i} \mathcal{F}_{m_1} \right) \left( \frac{1}{\mathcal{F}_i} + \sum_{m_2 \neq i} \frac{1}{\mathcal{F}_{m_2}} \right) \nonumber \\
    &= \Delta t^2 \left( 1 + a \mathcal{F}_i + b \frac{1}{\mathcal{F}_i} + ab \right) \label{eq:discretised_int}
\end{align}
with $a,b$ two positive coefficients independent of $\mathcal{F}_i$. For any $i=1,...,N=T/\Delta t$, Eq.~\eqref{eq:discretised_int} is maximised with respect to $\mathcal{F}_i$ when $\mathcal{F}_i = \mathcal{F}(0)=1$ or $\mathcal{F}_i = \mathcal{F}(T)=e^{2\Omega\lambda_\M/\gamma_\M}$. Combined with the fact that $\mathcal{F}$ is monotonically increasing, we conclude that $\Gamma(t)$ is a Heaviside step function and thus $\omega^*(t) \propto \delta(t-t_{\rm pulse})$. Extremising with respect to $t_{\rm pulse}$ finally gives Eq.~\eqref{eq:delta_optimum}.

\section{Total variation regularisation}\label{a:regularisation}
We can regularise the optimal protocol under the constraint of total shear by augmenting the action \eqref{eq:s_minimal} with a total variation regulator to penalise sharp changes in the shear rate,
\begin{equation}\label{eq:mS_regularised}
    \mS_{\rm reg}[\omega] = \mS[\omega] - 2 \mu \int_0^T dt (\omega'(t))^2\;, \quad \mu>0.
\end{equation}
For the sake of brevity, we only consider the limiting case of small $\lambda_\M$, for which the action is quadratic, Eq.~\eqref{eq:S_expanded_smalllambda}. Denoting by $G(t)$ the second antiderivative of the shear rate $\omega(t)$, $G(t)=\int_0^t dt' \int_{0}^{t'} dt'' \omega(t'') $, we obtain, after some integrations by parts and using $G(0)=0$,
\begin{equation}
    \mS_{\rm reg}[G] = 2\int_0^T ds (T-s)G''(s)(sG'(s)-G(s)) - \mu [G'''(s)]^2\;.
\end{equation}
We can now derive the associated Euler-Lagrange equation: $TG^{(2)}(t) - 2 \mu G^{(6)}(t)=0$. It is solved by the (smooth) function
\begin{equation}\label{eq:opt_regularised}
    \omega^*(t) = c_1 \cos(\alpha t) + c_2 \sin(\alpha t) + c_3 e^{\alpha t} + c_4 e^{-\alpha t}
\end{equation}
with $\alpha = (t/2\mu)^\frac{1}{4}$ and $c_i \in \mathbb{R}$. Rather than deriving particular expressions for $c_i(\Omega,\mu)$, we limit ourselves to pointing out that -- assuming symmetric boundary conditions $\omega^*(0)=\omega^*(T)$ -- the optimum will be time-reversal symmetric for all $\Omega$ and $\mu$. This follows from substituting \eqref{eq:opt_regularised} back into the action \eqref{eq:mS_regularised} and noticing that the resulting expression is a quadratic function of the integration coefficients. Since both the boundary conditions and the constraint of total rotation amount to linear relations amongst the $c_i$, the global maximum (obtained by extremising the action with respect to the one remaining free parameter) must be unique and thus symmetric. Note also that in the regularised problem we do not need to break the $\mathcal{P}$ symmetry explicitly for the optimum to be well-defined, since singular behaviour at the boundaries is sufficiently penalised by the regulator.

\section{Derivation of Eq.~\eqref{eq:ode_nonl}}\label{a:derivation_ODE}
Consider Eq.~\eqref{eq:diff_imp} with the functional derivative $\delta\mS/\delta\omega(t)$ as given in Eq.~\eqref{eq:S_func_der}. Again, let $\Gamma(t)$ denote the antiderivative of the rescaled shear rate, whereby $\int_{t_1}^{t_2}dt \lambda_\M\omega(t)=\Gamma(t_2)-\Gamma(t_1)$. Differentiating once, twice and three times with respect to $t$ we obtain, respectively,
\begingroup
\allowdisplaybreaks
\begin{align}
    \bar{\mu}_\sigma \omega'(t)&=\int_0^T dt_1 \sinh[2(\Gamma(t_1)-\Gamma(t))] \label{eq:d1}  \\
    \bar{\mu}_\sigma \omega''(t) &=-2\lambda_\M \omega(t)\int_0^T dt_1 \cosh[2(\Gamma(t_1)-\Gamma(t))] \label{eq:d2} \\
    \bar{\mu}_\sigma \omega'''(t)&=4\lambda_\M^2 \omega^2(t)\int_0^T dt_1 \sinh[2(\Gamma(t_1)-\Gamma(t))] \nonumber \\
    &\qquad  - 2\lambda_\M \omega'(t) \int_0^T dt_2 \cosh[2(\Gamma(t_1)-\Gamma(t))] \label{eq:d3}
\end{align}
\endgroup
with $\bar{\mu}_\sigma \equiv \mu_\sigma  \lambda_\M^2 /\gamma_\M^2$. Substituting Eqs.~\eqref{eq:d1} and \eqref{eq:d2} into the right-hand side of Eq.~\eqref{eq:d3}, we can eliminate the integrals and obtain the $\bar{\mu}_\sigma$-independent ODE
\begin{equation}
    \frac{\omega'''}{\omega} - \frac{\omega' \omega''}{\omega^2} - 4 \lambda_\M^2 \omega \omega' = \frac{d}{dt}\left( \frac{\omega''}{\omega} - 2 \lambda_\M^2 \omega^2\right) =0\;.
\end{equation}
Eq.~\eqref{eq:ode_nonl} follows by direct integration, with $c^2$ the integration constant.

\section{Looser bound on mixing efficiency at fixed dissipation}\label{a:loose_bound}

Except for the case $\lambda_\M=0$, for which the closed-form expression \eqref{eq:bound_Mdep_sigma} was obtained, we have seen in the main text that evaluating the tight bound on the mixing efficiency at fixed dissipation requires numerical integration as a final step.
Nevertheless, the action $\mS$ being a monotonically increasing function of $\omega(t)$ for any $t$, we may still construct an explicit, albeit looser, upper bound for $\lambda_\M>0$ by evaluating Eq.~\eqref{eq:s_minimal} at constant $\omega(t) = \omega_{\rm max}$ with $\omega_{\rm max} \equiv \max_\tau \omega^*_{\rm diss.}(\tau)$. From inspection of Eq.~\eqref{eq:sn_sol_sigmaparam} we find that
\begin{equation}
    \omega_{\rm max} = \frac{2 K(m)\sqrt{m}}{\lambda_\M T}~,
\end{equation}
where we recall that the parameter $0 \leq m < 1$ is fixed by the implicit equation $4K(m)(K(m)-E(m)) = \sigma_\M \lambda_\M^2 T$. The following bound on the mixing efficiency follows by direct integration of the action functional:
\begin{align}
    \left. \Delta I(\sigma;M) \right|_{\lambda_\M > 0} \leq \frac{1}{2} \ln \left[ 1 + \frac{\gamma_\M^2 }{4 \xi_\M^2} \left(\sinh^2(\xi_M) - \xi_M^2\right) \right]
    \label{eq:bound_Mdep_sigma_approx} 
\end{align}
where $\xi_\M \equiv 2 \lambda_\M \omega_{\rm max} T$. 
Since $\lim_{\sigma \lambda_\M^2 T \to \infty} m K^2(m) =\sigma_\M \lambda_\M^2 T/4$, which follows from the implicit equation for $m$, the right-hand side of Eq.~\eqref{eq:bound_Mdep_sigma_approx} has a square root asymptote in the limit of large dissipation, specifically
\begin{equation}\label{eq:loose_sqrt_limit}
    \left.\Delta I(\sigma;M)\right|_{\lambda_\M > 0} \leq \sqrt{\frac{4 \sigma \lambda_\M^2 T}{\eta \gamma_\M^2}}
\end{equation}  
for $\sigma_\M \lambda_\M^2 T \gg 1$. We conclude that, for any $\M$ with $\lambda_\M >0$, the tight bound \eqref{eq:bound_Mdep_sigma} cannot grow faster than $\sqrt{\sigma}$ in the large dissipation regime. Indeed, numerical evidence indicates that this is also the asymptotic scaling of the tight bound for pure shear.

\end{document}